\pdfoutput=1

\documentclass[11pt]{article}

\usepackage[final]{acl}

\usepackage{times}
\usepackage{latexsym}

\usepackage[T1]{fontenc}

\usepackage[utf8]{inputenc}

\usepackage{microtype}

\usepackage{inconsolata}

\usepackage{graphicx}

\usepackage{hyperref}       
\usepackage{url}            
\usepackage{booktabs}       
\usepackage{amsfonts}       
\usepackage{nicefrac}       
\usepackage{microtype}      
\usepackage{xcolor}         
\usepackage{adjustbox}
\usepackage{graphicx}
\usepackage{subcaption}
\usepackage{array}
\usepackage{amsmath}
\usepackage{enumitem}
\usepackage{algorithm}
\usepackage{algpseudocode}
\usepackage{multirow}
\usepackage{float}
\usepackage{tabularx}
\usepackage{colortbl}
\usepackage{makecell}
\usepackage{wrapfig}
\usepackage{pifont}

\title{MolecularGPT: Open Large Language Model (LLM) \\for Few-Shot Molecular Property Prediction}

\author{Yuyan Liu$^{1,2}$, Sirui Ding$^3$, Sheng Zhou$^4$, Wenqi Fan$^5$, Qiaoyu Tan$^1$ \\
  $^1$ New York University Shanghai \\
  $^2$ Xiamen University, $^3$ University of California, San Francisco \\
  $^4$ Zhejiang University, $^5$ The Hong Kong Polytechnic University \\
  \texttt{yanyanyan@stu.xmu.edu.cn}, \texttt{qiaoyu.tan@nyu.edu} \\}

\begin{document}
\maketitle
\begin{abstract}
Molecular property prediction (MPP) is a fundamental and crucial task in drug discovery. However, prior methods are limited by the requirement for a large number of labeled molecules and their restricted ability to generalize for unseen and new tasks, both of which are essential for real-world applications. To address these challenges, we present MolecularGPT for few-shot MPP. From a perspective on instruction tuning, we fine-tune large language models (LLMs) based on curated molecular instructions spanning over 1000 property prediction tasks. This enables building a versatile and specialized LLM that can be adapted to novel MPP tasks without any fine-tuning through zero- and few-shot in-context learning (ICL). MolecularGPT exhibits competitive in-context reasoning capabilities across 10 downstream evaluation datasets, setting new benchmarks for few-shot molecular prediction tasks. More importantly, with just two-shot examples, MolecularGPT can outperform standard supervised graph neural network methods on 4 out of 7 datasets. It also excels state-of-the-art LLM baselines by up to 15.7\% increase on classification accuracy and decrease of 17.9 on regression metrics (e.g., RMSE) under zero-shot. This study demonstrates the potential of LLMs as effective few-shot molecular property predictors. The code is available at \url{https://github.com/NYUSHCS/MolecularGPT}.

\end{abstract}

\section{Introduction}

The discovery of molecules with desired functional properties is crucial for advancements in fields such as medicine~\citep{cell,nature,science,AlphaFold} and material~\citep{materials,mof}. Molecular property prediction (MPP), which employs deep learning techniques to predict molecules' functional properties, has proven effective in accelerating the drug discovery process and reducing associated costs~\citep{nature,materials,mof}. 

Among them, graph neural networks (GNNs)-based methods~\citep{GAT,GIN,GCN,mpnn,graphsage} have achieved state-of-the-art results in the past few years. However, these methods~\citep{GeomGCL,graphmvp,Infomax} are limited in supervised settings, contradicting with practical needs as annotating molecules is both expensive and time-consuming. Furthermore, the task-specific supervised learning process may hurdle the model's adaptation to new tasks, limiting its generalization ability in open-world scenarios.

Inspired by this, several recent endeavors have aimed to enable zero-shot reasoning for MPP~\citep{clamp, GIMLET} by integrating both natural language and molecular representations. CLAMP~\citep{clamp} is a text-molecule model that aligns pairs of chemical text (e.g., descriptions of molecular properties) and molecule graphs through contrastive learning. Subsequently, the bioactivity of a query molecule is classified by measuring the similarity between its molecular representation and corresponding bioassay description. While effective, CLAMP is limited to classification tasks and is not a generative model. 

In contrast, another line of research in LLMs~\citep{GIMLET} integrates molecule graphs and task descriptions into a unified generative LLM. This approach enables zero-shot reasoning for molecular property prediction across both classification and regression tasks. However, the inclusion of an additional architectural design restricts it from performing few-shot molecular property predictions, a capability naturally supported by standard LLMs. 

To date, there's no LLM-based method in the molecular domain fully inherits the generalization and ICL abilities of LLMs as seen in the NLP field, which raises a research question: 
\textit{Can LLMs be fine-tuned for generic MPP, enabling the resultant model to generalize to a variety of unseen tasks and inherit LLMs' few-shot ICL ability? }


In this work, we aim to bridge the gap and present MolecularGPT, the first instructionally tuned LLM that can generalize to a variety of novel MPP tasks while retaining its zero-shot and few-shot in-context reasoning abilities. Specifically, MolecularGPT adopts the SMILES~\citep{smiles} representation of molecules as a unified graph-to-string transformation for instruction construction, as it precisely translates molecules' chemical structures into a string of atomic symbols and chemical bonds based on a set of rules. To fully utilize the graph structures in molecules, we introduce structure-aware few-shot instructions, which incorporate the top-$k$ neighbors, globally retrieved based on their similarities, of each molecule as complementary information for instruction design. This design aligns the instruction tuning and inference prompt format of MolecularGPT, making it naturally applicable for few-shot ICL. Additionally, to balance zero-shot and few-shot reasoning capabilities, we explore various combination options and empirically find that a hybrid instruction set, including both zero-shot and few-shot instructions, enables MolecularGPT to perform well in both zero-shot and few-shot property predictions. Our \textbf{main contributions} are summarized below:

\begin{itemize}[leftmargin=*]
\item We study how to adapt pre-trained LLMs to molecular field, enabling effective few-shot MPP in the ICL fashion. Specifically, we propose MolecularGPT, the first instructionally fine-tuned LLM that supports few-shot property prediction on unseen tasks without any fine-tuning. 

\item We introduce the concept of structure-aware few-shot instruction to better adapt LLMs with molecular field. Unlike existing efforts~\citep{clamp,GIMLET,moleculegpt} that focus on fusing graph structures and SMILES representations in a model-centric perspective, we maliciously combine them in a data-centric manner by constructing global structure-aware few-shot demonstrations.




\item We devise a hybrid instruction set to inherit the few-shot ICL capability of LLMs. This set is a mix of both few-shot and zero-shot instructions that span over 1000 MPP tasks including both classification and regression tasks across biological, chemical, and quantum mechanical domains, resulting in 3.5GB training tokens. This diversified instruction set has been empirically proved to be effective in adapting LLMs for MPP tasks.  

\item We extensively experimented on 10 molecular property benchmarks across different scales and tasks to validate the effectiveness of MoelcularGPT. Our empirical results demonstrate that MoelcularGPT outperforms the leading LLM baselines (e.g., GIMLET and LLaMA-7b~\citep{llama}), with up to an average 15.7\% improvement across all classification tasks. Additionally, with just two-shot examples, MolecularGPT surpass standard supervised GNN methods on 4 out of 7 datasets, setting new benchmarks for few-shot molecular property tasks.
\end{itemize}


\begin{figure*}[!t]
\centering
\includegraphics[width=1\linewidth]{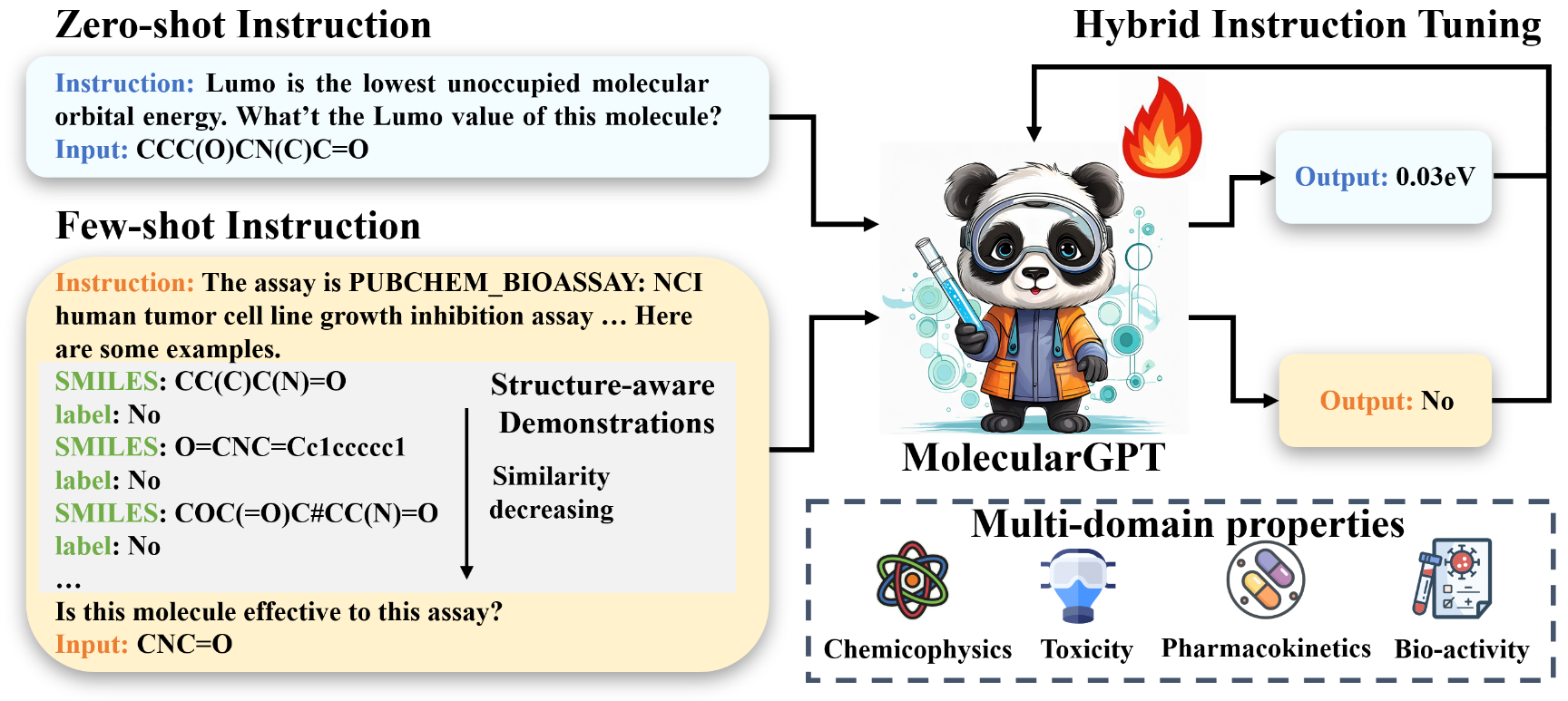}
\caption{The proposed MolecularGPT framework. To instructionally fine-tune LLMs for MPP tasks, we construct a hybrid instruction set that includes both zero-shot and few-shot instructions across more than 1000 property tasks. Each few-shot instruction adaptively selects the query molecule's top-$k$ neighboring molecules as labeled demonstrations for prompt design.}
\label{fig:map}
\vspace{-0.6 cm}
\end{figure*}


\section{Related work}

\noindent\textbf{GNNs-based MMP}
GNNs~\citep{GAT,GIN,GCN,Infomax} perform MPP tasks by constructing models between molecular graphs and properties. {Though these models have achieved great success~\citep{mpnn,graphsage,GeomGCL,graphmvp,tan2023s2gae}, they are implicitly trained for each task without explicit natural language instructions, limiting their ability to generalize to new tasks. While a few approaches~\citep{guo2021few,wang2021property,schimunek2023context,zhuang2023graph} aim to extend pre-trained GNNs to new tasks, they require a large number of related tasks to pre-train a meta-learner, making them label-intensive and deviating from our generalization goals.} Moreover, these supervised models rely solely on structure information, overlooking the rich textual knowledge derived from wet lab experiments.

\noindent\textbf{Pretrain-finetune based molecular language models}
To utilize the chemical knowledge in texts, molecular language models~\citep{moleculeSTM,molt5,biot5,moleculegpt,molca} aim to integrate natural language and molecular representations for joint reasoning. These models~\citep{momu,kvplm,MolXPT,gitmol,3D-MoLM} involve two stages: pre-training and fine-tuning. The pre-training phase primarily focuses on learning molecular representations and their associated textual descriptions through masked language modeling, contrastive learning or next token prediction. However, they still require fine-tuning on particular MPP downstream tasks, thereby limiting their generalization abilities to new tasks.

\noindent\textbf{Instruction tuning based molecular language models}
To address this, recent efforts in molecular language modeling~\citep{mol-instructions,GIMLET} aim to explicitly align molecular graphs with their properties through instruction tuning \cite{Collection}. For instance, GIMLET~\citep{GIMLET} integrates molecular graphs with instruction languages for fine-tuning LLMs. GIMLET achieves effective zero-shot ICL for new tasks but lacks few-shot ICL capability due to its generalized position embedding and decoupled attention designs. Mol-Instructions~\citep{mol-instructions} is a close work to us, but it fine-tunes LLMs with only three properties tasks and neglects inter-molecular correlations, significantly limiting its zero-shot and few-shot ICL performances. In contrast, we curate a diverse instruction set covering 1000 property tasks and introduce structure-aware few-shot instructions to significantly enhance the zero-shot and few-shot reasoning capabilities of LLMs in MPP tasks. More details about these property tasks can be found in Appendix~\ref{sub:Details of datasets}



\section{Method}

In this section, we present the proposed MolecularGPT, as shown in Fig.~\ref{fig:map}. First, we discuss the general instructional fine-tuning pipeline to adapt LLMs for MPP tasks (in Section~\ref{sec-it}). Next, we elaborate on a structure-aware few-shot instruction design strategy to effectively incorporate graph structures among molecules (in Section~\ref{sec-fsit}). Finally, we illustrate a hybrid instruction tuning approach that enhances both the zero-shot and few-shot reasoning capabilities of LLMs for MPP tasks (in Section~\ref{sec-hist}). 

\noindent\textbf{Notations and Problem Formulation.} Given a set of $n$ molecular graphs $D=\{(G_i, y_i)|i\in 1,2,...,n\}$, where $G_i=(\mathcal{V}, \mathcal{E})$ represents the $i$-th molecule and $y_i$ is the ground-truth property (e.g., categorical label or numerical score). Here, $\mathcal{V}$ and $\mathcal{E}$ denote the node set and edge set, respectively. The goal of molecular instruction tuning is to fine-tune a LLM model $f_\theta$ by fitting a set of training instructions $S_D$ (i.e., (\textit{input}, \textit{output}) pairs) constructed from $D$, so that the fine-tuned LLM can be directly applied to make property predictions for unseen tasks or molecules, i.e., $D_\text{test}=\{(G_j,y_j)|j=1,2,...,m\}$ with $D\cap D_\text{test}=\emptyset$. 

While conceptually simple, successfully achieving molecular instruction tuning involves addressing several research challenges. \textbf{C1}: how can we unify molecules of varying sizes, densities, and domains into a consistent format, ensuring that important molecular information in $D$ and $D_\text{test}$ is consistently incorporated? \textbf{C2}: given that graph structures are crucial for molecular analysis, as verified in GNN studies, how can we effectively include these structures in molecular instruction tuning? \textbf{C3}: considering that molecule annotation is notoriously expensive and time-consuming, how can we enable the fine-tuned LLM to benefit from few-shot scenarios where only a few labeled molecules are available in real-world applications?  

\subsection{SMILES-based Molecular Instruction Tuning: A Unified Step}
\label{sec-it}
{To improve the generalization capability of fine-tuned LLM for MPP tasks (\textbf{C1}), 
we aim to employ the well-known graph-derived linear strings~\citep{selfies} of molecular graphs, SMILES~\citep{smiles} for instruction tuning. Different from GNN encoders~\citep{clamp,GIMLET}, SMILES translates molecules' chemical structure into a string of atomic symbols and chemical bonds based on a set of rules~\citep{qian2023can}. This precise translation provides a universal expression foundation for different types of molecules. Following standard instruction tuning protocol~\citep{TextChem,moleculegpt,gitmol,3D-MoLM}, the molecular instruction set $S_D$ can be generated by the following prompt template $T=\{Q, I, R\}$ based on $D$, regarding as a zero-shot instruction template.}

\begin{quote}
\label{templte:template}
\begin{verbatim}
### Instruction: {instruction} 
### Input: {inputs} 
### Response: {output}.
\end{verbatim}
\end{quote}

Here, the instruction question $Q$, SMILES strings of query molecule $I$, and property label $R$ are mapped to the \{instruction\}, \{inputs\}, and \{output\} components, respectively.

\subsection{Structure-Aware Molecular Instruction Tuning: Graph Structure Matters}
\label{sec-fsit}
So far, we have illustrated how to incorporate graph structure within each molecule into instruction via the zero-shot instruction template $T$. However, this approach may result in subpar prediction performance due to the neglect of correlations between molecules. To address this, we introduce structure-aware instruction tuning (\textbf{C2}), which aims to incorporate inter-molecular structures into the prompt template. The high-level idea is to utilize similar molecules as demonstrations to enhance LLM reasoning. 

To achieve this, given a query molecule $G_i\in D$, we identify its top-$K$ nearest molecules in $D$ based on the following retrieval module.    
\begin{equation}
N_{G_i} = \text{topK}(G_i, D, K),
\end{equation}
where $N_{G_i}$ is the retrieved neighborhood set with $K$ molecules. $\text{topK}()$ is a search algorithm based on the similarity between molecules. Specifically, we estimate the similarity between molecules by calculating their Tanimoto coefficient~\citep{tanimoto} based on their MACCS Keys~\citep{maccs}. Notably, MACCS Keys, comprising 166 binary keybits, provides a unified representation for molecules and has been widely adopted in many molecule retrieval systems, such as USearch~\citep{USearch}.  

Utilizing $N_{G_i}$, we can transform the zero-shot template into a few-shot version $T_{shot}=\{C, I, R\}$, where $C$ represents the k-shot instruction question, extending $Q$ with structurally similar molecule demonstrations extracted from $N_{G_i}$. Specifically, let $(\textit {m}_{i}, \textit {y}_{i})$ represents the $\textit {i}$-th similar molecule-property pairs in $N_{G_i}$. Additionally, considering that the order of demonstrations may significant impact prompt design~\citep{mosbach2023few}, we arrange these k demonstrations in a descending order based on their similarity scores. The $C$ is formally expressed as:
\begin{equation}
\label{fun:fun}
\textit {C} = \{\textit {Q}, ( (\textit {m}_{1}, \textit {y}_{1}),..., (\textit {m}_{i}, \textit {y}_{i}),...,(\textit {m}_{k},\textit {y}_{k})) \}.
\end{equation}

Similar to $T$, the extended question $C$ in the few-shot instruction template $T_{shot}$ will correspond to the \{instruction\} of the template in Section~\ref{templte:template}. In experiments, we empirically observed that including the target property of molecular neighbors as input in few-shot scenarios improves performance.
This approach is reasonable because $T_{shot}$ serves as a few-shot in-context prompt, akin to those widely used in the NLP domain, where the most similar neighbors are selected as demonstrations.

\subsection{Hybrid Molecular Instruction Tuning: Better Few-Shot Learner}
\label{sec-hist}

Given the advanced structure-aware instruction template $T_{shot}$, one can easily construct the instruction training set $S_D$ by applying $T_{shot}$ on each molecule in $D$. Then, we fine-tune a pre-trained LLM by optimizing the following training loss: 
\begin{equation}
\mathcal{L}(\theta) = \sum_{(C_i, I_i, R_i)) \in \mathcal{S}_D} -\log f_{\theta}(R_i|C_i, I_i).
\end{equation}
Here, $f_\theta$ is a pre-trained LLM with parameter $\theta$. In practice, we initialize $f_\theta$ as LLaMA2-7b-chat~\citep{llama} and adopt QLoRA~\citep{qlora} to speedup the training. 

\begin{figure}[!t]
\centering
\includegraphics[width=1\linewidth]{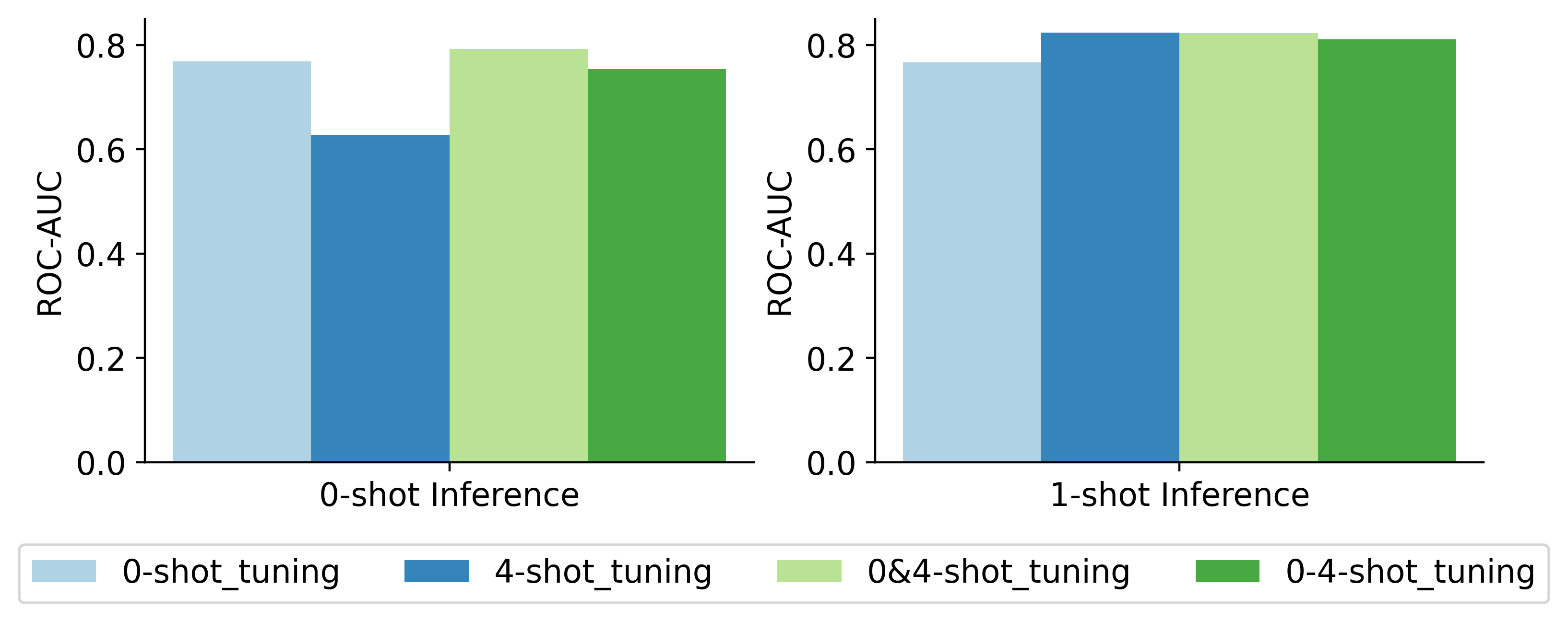}
\caption{The performance on Cyp450 test dataset.}
\label{fig:small}
\vspace{-0.5cm}
\end{figure}

While $T_{shot}$ appears effective, it may degrade the zero-shot reasoning capability of fine-tuned LLM due to the explicit graph structures among molecules. To verify this, we conducted a toy example by fine-tuning LLaMA2-7b-chat on different $K$-shot instruction sets. Specifically, Fig.~\ref{fig:small} reports the zero-shot and one-shot inference results on the CYP450 dataset for $K=0$ and $K=4$. 

In Fig.~\ref{fig:small}, we can observe an obvious trade-off between zero-shot and one-shot performance with respect to the instruction set. For example, when fine-tuning LLaMA2 on the 0-shot instruction set constructed using the $T$ template, the resulting \textit{0-shot\_tuning} model performs well in zero-shot scenarios but underperforms in one-shot scenarios. Conversely, when fine-tuning on the 4-shot instruction set constructed using $T_{shot}$ with $K=4$, the resulting \textit{4-shot\_tuning} model excels in one-shot settings but underperforms in zero-shot cases.

This observation motivates us to introduce a \textbf{hybrid instruction set} $S_D^h$, combining the strengths of both the zero-shot instruction template $T$ and the few-shot instruction template $T_{shot}$. Specifically, $S_D^h$ is derived from a combination of 0, 1, 2, 3, and 4-shot instruction templates. In Fig.~\ref{fig:small}, we can see that our hybrid instruction tuned models, \textit{0\&4-shot\_tuning} and \textit{0-4-shot\_tuning}, consistently outperforms others in zero-shot and one-shot scenarios. Further details can be found in Section~\ref{sub:mixture}.

\section{Experiment}

\begin{table*}
\vspace{-0.5cm}
\scriptsize
\renewcommand\arraystretch{0.5}
  \caption{Performance over Bio-activity (Bio), Toxicity (Tox), and Pharmacokinetic (Pha) classification tasks. Highlights are the {\color{red}\textbf{first}} and {\color{blue}\textbf{second}} best results over all 2-shot and 8-shot performances. We also report the average rank (Avg.RK) of all models under 2-shot and 8-shot separately.
  }
  
  \centering
  \resizebox{\textwidth}{!}{
  \begin{tabular}{llccccccccccc>{\columncolor{gray!20}}c}
    \toprule
       \textbf{Type} & \textbf{Method} & \textbf{Shot} & \textbf{BACE} & \textbf{HIV} & \textbf{MUV} & \textbf{Avg.Bio} & \textbf{Tox21} & \textbf{ToxCast} & \textbf{Avg.Tox}  & \textbf{BBBP} & \textbf{CYP450} & \textbf{Avg.Pha} & \textbf{Avg.RK} \\\\
    \midrule
    \multirow{14}{*}{\makecell[c]{Few-shot \\ Fintuned}} & \multirow{2}{*}{Pre-GS-Meta} & 2-shot & 0.5213 & 0.5702 & 0.5507  & 0.5474 & 0.5976 &  0.5310 & 0.5643 & 0.6668 & 0.5105  & 0.5887 & 6.1\\
    && 8-shot & 0.6203 & 0.6128 & 0.5791 & 0.5474 & 0.5665 & 0.5208  & 0.5437 & 0.6004 & 0.6378 & 0.6191 & 6.4\\
    \cmidrule(r){2-14}
    &\multirow{2}{*}{KVPLM} & 2-shot & 0.6380  & 0.6675 & 0.5296  & 0.6117 & 0.5336 & 0.5122  & 0.5229 & 0.5399 & 0.5960  & 0.5680 & 6.0\\
    && 8-shot & 0.6375 & 0.6293 & 0.4987 & 0.5885 & 0.5712 & 0.4892 & 0.5302 & 0.5401 & 0.6305 & 0.5853 & 7.0\\
    \cmidrule(r){2-14}
    &\multirow{2}{*}{MoMu} & 2-shot & 0.5747  & 0.6435 & 0.4701   & 0.5568 & 0.5297 & 0.5164  & 0.5231 & 0.5901 & 0.5976  & 0.5939 & 6.7\\
    && 8-shot & 0.6102 & 0.6171 & 0.5804 & 0.6026 & 0.5517 & 0.5226 & 0.5372 & 0.5728 & 0.5713 & 0.5721 & 7.0\\
    \cmidrule(r){2-14}
    &\multirow{2}{*}{GIMLET} & 2-shot & 0.7141  & 0.6794 & 0.6420 & 0.6785 & 0.6222 & 0.5903  & 0.6063 & 0.6094 & 0.7212  & 0.6653 & 2.4 \\
    && 8-shot & 0.7081 & 0.6803 & {\color{red}\textbf{0.6487}} & {\color{blue}\textbf{0.6790}} & 0.6197 & 0.5911 & 0.6054 & 0.6119 & 0.7234 & 0.6677 & 2.9\\
    \midrule
    \multirow{18}{*}{\makecell[c]{Few-shot \\ Inference}} & \multirow{2}{*}{MHNfs} &  2-shot & 0.5939 & 0.5835 & 0.5566 & 0.5780 & 0.5360 & 0.5370 & 0.5365 & 0.5750 & 0.5734  & 0.5742 & 6.0\\
    && 8-shot & 0.6294 & 0.6397 & 0.6764 & 0.6485 & 0.6162 & 0.5602 & 0.5892 & 0.5442 & 0.5992 & 0.5717 & 4.9\\
    \cmidrule(r){2-14}
    &\multirow{2}{*}{LLaMA2-7B} &2-shot & 0.6930 & 0.6587 & 0.5085 & 0.6201 & 0.6052 & 0.5010 & 0.5531 & 0.5459 & 0.5807  & 0.5633 & 5.9 \\
    && 8-shot & {\color{blue}\textbf{0.7271}} & {\color{blue}\textbf{0.7000}} & 0.5161 & 0.6477 & 0.6505 & 0.5329 & 0.5917 & 0.5127 & 0.6891 & 0.6009 & 4.0\\
    \cmidrule(r){2-14}
    &\multirow{2}{*}{Galactica125M} &2-shot & 0.6298  & 0.6089 & 0.4831   & 0.5739 & 0.5902 & 0.5090  & 0.5496 & 0.3931 & 0.6961  & 0.5446 & 6.7\\
    && 8-shot & 0.5248 & 0.6474 & 0.5302 & 0.5675 & 0.6083 & 0.5035 & 0.5559 & 0.4174 & 0.6217 & 0.5196 & 7.0\\
    \cmidrule(r){2-14}
    &\multirow{2}{*}{Galactica1.3B} &2-shot & 0.6634  & 0.5295 & 0.5171   & 0.5700 & 0.6302 & 0.5566  & 0.5934 & 0.6970 & 0.7323  & 0.7147 & 4.0 \\
    && 8-shot & 0.7152 & 0.6623 & 0.4977 & 0.6251 & 0.6470 & 0.5279 & 0.5875 & {\color{blue}\textbf{0.7092}} & 0.7645 & 0.7369 &3.7 \\
    \cmidrule(r){2-14}
    &\multirow{2}{*}{MolecularGPT} &2-shot & 0.7218 & {\color{red}\textbf{0.7204}} & 0.6338 & {\color{red}\textbf{0.6920}} & {\color{red}\textbf{0.6573}}  & {\color{blue}\textbf{0.5945}} & {\color{blue}\textbf{0.6259}} & {\color{red}\textbf{0.7260}} & {\color{red}\textbf{0.8275}} & {\color{red}\textbf{0.7768}} & \textbf{1.1} \\
    && 8-shot & {\color{red}\textbf{0.7331}} & 0.6382 & {\color{blue}\textbf{0.6469}} & 0.6727 & {\color{blue}\textbf{0.6565}} & {\color{red}\textbf{0.5985}} & {\color{red}\textbf{0.6275}} & 0.6822 & {\color{blue}\textbf{0.8228}} & {\color{blue}\textbf{0.7525}} & \textbf{2.1} \\
    \bottomrule
  \end{tabular}
  }
  \label{table1}
  \vspace{-0.5cm}
\end{table*}

In our experimental framework, we aim to answer three primary research questions: \textbf{RQ1:} Can MolecularGPT effectively handle new property prediction tasks through few- and zero-shot ICL? \textbf{RQ2:} What is the optimal design for in-context instruction set to improve MolecularGPT's generalization and ICL abilities during tuning? \textbf{RQ3:} How does the question prompts and retrieval in-context examples affect MolecularGPT's performance?

\subsection{Experimental Setup}

\paragraph{Datastes}
Consistent with the GIMLET setting, we employ the MoleculeNet benchmark~\citep{molecule-net} and CYP450~\citep{CYP450} datasets as our downstream datasets, totally 657 MMP tasks. More details about datasets can be found in Appendix~\ref{sub:Details of datasets}. We employ ROC-AUC as the evaluation metric for classification tasks, while the Root Mean Square Error (RMSE) for regression tasks. 

\paragraph{Baselines}
{We compare our MolecularGPT with 5 leading GNN-based methods: GCN~\citep{GCN}, GAT~\citep{GAT}, GIN~\citep{GIN}, Graphormer~\citep{graphomer}, and Graphormer-p~\citep{graphomer}, 5 SOTA LM-based molecular models: XVPLM~\citep{kvplm}, MoMu~\citep{momu}, Galactica-125M~\citep{galactica}, Galactica-1.3B~\citep{galactica}, and GIMLET~\citep{GIMLET}, and 2 strong few-shot molecular property prediction baselines: Pre-GS-Meta~\citep{Pre-GS-Meta} and MHNfs~\citep{MHNfs}, as well as the open-sourced LLM: LLaMA-chat-7B~\citep{llama}}

\subsection{Performance Evaluation}


As the results presented in Tab.~\ref{table1},~\ref{table2} respectively, MolecularGPT can achieve competitive performance on classification and regression tasks under both few-shot and zero-shot settings. We answer the \textbf{RQ1} with more details as follows.

\ding{172} \textbf{MolecularGPT establishes a new benchmark in few-shot learning.} As shown in Tab.~\ref{table1}, our model delivers the best performance across all datasets under the few-shot setting, securing the top-1 average rank across all datasets in both the 2-shot and 8-shot scenarios, with respective ranks of 1.1 and 2.1. Most notably, unlike models in the few-shot finetuned category, such as GIMLET, which require additional tuning and parameter updates, our MolecularGPT is capable of directly inference with few-shot examples, demonstrating an average increase of 4.3\% in 2-shot and 2.8\% in 8-shot scenarios respectively. Comparing with strongest model in few-shot inference category, Galactica1.3B, our model shows an average increase of 7.9\% in 2-shot and 3.6\% in 8-shot scenarios separately, proving its strongest generalization capabilities. Additionally, MolecularGPT even surpasses supervised finetuned GNNs in 4 out of 7 classification tasks in a 2-shot setting, as shown in Tab.~\ref{table1} and \ref{table2}.



\ding{173} \textbf{MolecularGPT exhibits exceptional performance in zero-shot learning.} As shown in Tab. \ref{table2}, during zero-shot inference, our model achieves the top-1 average rank across all datasets, outperforming the strongest baseline, GIMLET, on 5 out of 10 datasets. It demonstrates an average improvement of 6.2\% on 3 classification tasks and an average decrease of 0.16 on 2 regression tasks. Comparing with our base model LLaMA, our model shows an average 15.7\% improvement on classification tasks and 17.9 decrease on regression tasks. Furthermore, when compared with GNNs that have been supervised finetuned on complete training set of each dataset, our model demonstrates comparable performance on the HIV, BBBP, CYP450, and ESOL datasets, highlighting its efficacy in a zero-shot setting.

\begin{table*}
\scriptsize
\vspace{-0.5cm}
\renewcommand\arraystretch{0.5}
  \caption{Performance over Bio-activity, Toxicity, and Pharmacokinetic classification tasks as well as Physicalchemical regression tasks. Highlights are the {\color{red}\textbf{first}} and {\color{blue}\textbf{second}} best results of 0-shot performances. In supervised finetuned models, we also mark the \textbf{highest} and \underline{\textbf{lowest}} results. Here, "finetuned" refers to the model being trained on the complete training set.}
  
  \centering
  \resizebox{\textwidth}{!}{
  \begin{tabular}{lccccccccccc>{\columncolor{gray!20}}c}
    \toprule
    \textbf{Method}  & \textbf{Type} & \textbf{BACE} & \textbf{HIV} & \textbf{MUV} &  \textbf{Tox21} & \textbf{ToxCast} & \textbf{BBBP} & \textbf{CYP450} & \textbf{ESOL} & \textbf{FreeSolv} & \textbf{Lipo} & \textbf{Avg.RK} \\\\
    \midrule
    XVPLM  & \multirow{16}{*}{\makecell[c]{0-shot \\ Inference}}& 0.5126 & 0.6120 & 0.6172  & 0.4917 & 0.5096   & 0.6020 & 0.5922 & - & - & - & 4.3 \\\\
    MoMu  & & {\color{blue}\textbf{0.6656}} & 0.5026 & 0.6051  & 0.5757 & 0.5238   & 0.4981 & 0.5798 & - & - & - & 3.9 \\\\
    Galactica125M & & 0.4451 & 0.3671 & 0.4986  & 0.4964 & 0.5106  & {\color{blue}\textbf{0.6052}} & 0.5369 & 2.130 & {\color{red}\textbf{3.660}} & {\color{blue}\textbf{1.097}} & 4.4 \\\\
    Galactica1.3B & & 0.5648 & 0.3385 & 0.5715  & 0.4946 & 0.5123  & 0.5394 & 0.4686 & {\color{red}\textbf{1.103}} & {\color{blue}\textbf{4.267}} & {\color{red}\textbf{1.093}} & 4.1 \\\\
    LLaMA2-7B & & 0.4911 & 0.6060 & 0.5554  & 0.5481 & 0.4693  & 0.3671 & 0.4198 & 7.227 & 15.912 & 2.329 & 5.6 \\\\
    GIMLET &  & {\color{red}\textbf{0.6957}} & {\color{blue}\textbf{0.6624}} & {\color{red}\textbf{0.6439}}  & {\color{red}\textbf{0.6119}} & {\color{red}\textbf{0.5904}}  & 0.5939 & {\color{blue}\textbf{0.7125}} & {\color{blue}\textbf{1.132}} & 5.103 & 1.345 & 2.2 \\\\
    MolecularGPT & & 0.6212 & {\color{red}\textbf{0.7128}} & {\color{blue}\textbf{0.6253}} & {\color{blue}\textbf{0.5893}} & {\color{blue}\textbf{0.5669}} & {\color{red}\textbf{0.6373}} & {\color{red}\textbf{0.8031}} & 1.471 & 4.975 & 1.157 & \textbf{2.1} \\\\
    \midrule
    GCN & & 0.736 & 0.757 & 0.732  & 0.749 & \underline{\textbf{0.633}} & \underline{\textbf{0.649}} & \underline{\textbf{0.8041}} & \underline{\textbf{1.331}} &  2.119 &   0.760 & 3.6\\\\
    GAT &  & \underline{\textbf{0.697}} & \underline{\textbf{0.729}} & \underline{\textbf{0.666}} &  0.754 & 0.646   & 0.662 & 0.8281  & 1.253 & 2.493 & 0.770 & 4.3 \\\\
    GIN & Finetuned & 0.701 & 0.753 & 0.718   & \underline{\textbf{0.740}} & 0.634  & 0.658 & 0.8205 & 1.243 & \underline{\textbf{2.871}} & \underline{\textbf{0.781}} & 4.0 \\\\
    Graphormer & & 0.7760 & 0.7452 & 0.7061  & 0.7589 & 0.6470  & 0.7015 & 0.8436 & 0.901 & 2.210 & 0.740 & 2.5 \\\\
    Graphormer-p & & \textbf{0.8575} & \textbf{0.7788} & \textbf{0.7480}  &  \textbf{0.7729} & \textbf{0.6649}  & \textbf{0.7163} & \textbf{0.8877} & \textbf{0.804} & \textbf{1.850} & \textbf{0.675}  & \textbf{1.0} \\\\
    \bottomrule
  \end{tabular}
  }
  \label{table2}
\end{table*}




\subsection{Tuning on Hybrid Instruction Set}
\label{sub:mixture}

\begin{figure*}[!t]
\centering
\vspace{-0.3cm}
\includegraphics[width=1\linewidth]{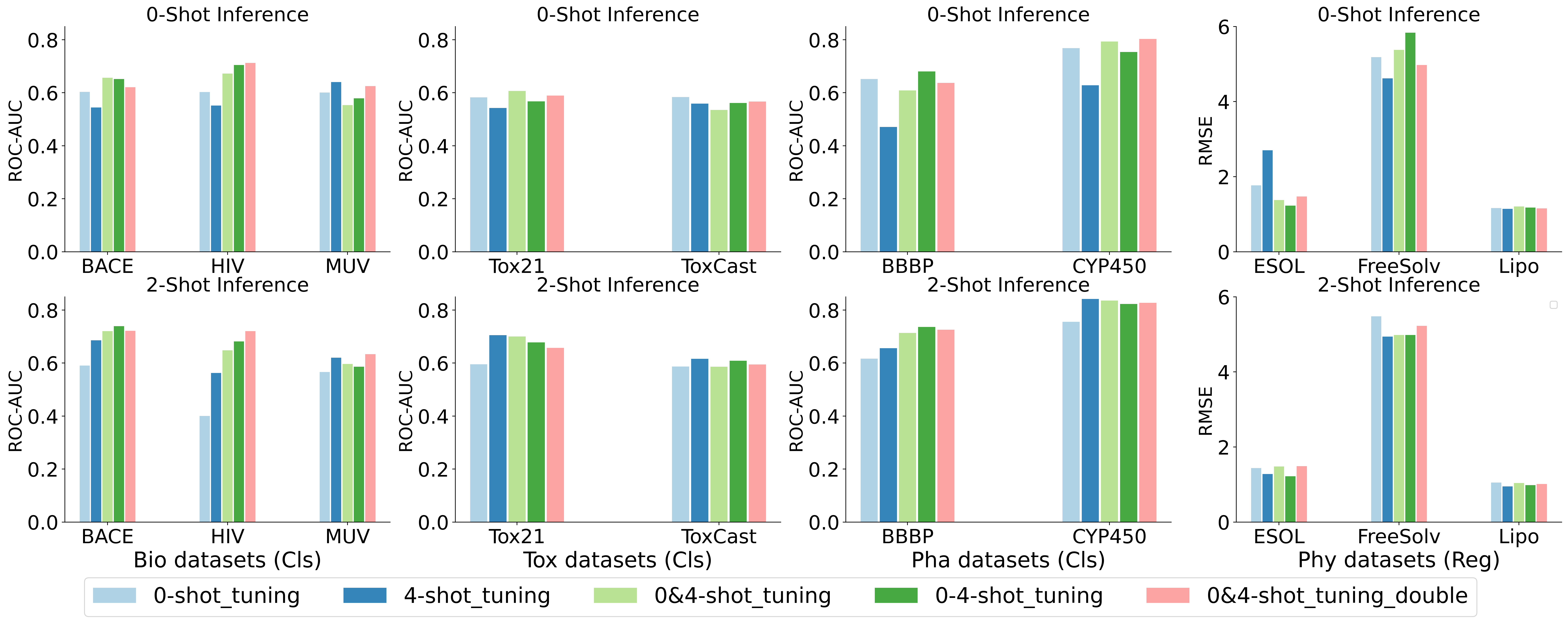}
\caption{The performance of MolecularGPT on Classifcation (Cls) and Regression (Reg) tasks tuning with different types of instruction datasets. We inference them with 0-shot and 2-shot examples. (\textit{0\&4-shot} indicates hybrid of $0$ and $4$-shot. \textit{0-4-shot} indicates mix of 0,1,2,3,4-shot. \textit{tuning\_double} indicates double the instruction set size.)}
\label{fig:performance}
\vspace{-0.5cm}
\end{figure*}

To investigate the \textbf{RQ2}, we conduct experiments to study the effect of hybrid instruction tuning set as presented in Fig.~\ref{fig:performance}.



\ding{174} \textbf{Tuning on property descriptions without demonstrations can improve the zero-shot performance.} As shown in the \textit{0-shot\_tuning} in Fig.~\ref{fig:performance}, the model performed satisfactorily on some tasks under 0-shot inference but poorly on many tasks under 2-shot inference. In comparison with 2-shot inference, 0-shot inference even demonstrates an average improvement of 4\% on classification tasks. We speculated that the zero-shot instruction set imparts some knowledge to LLaMA without significantly enhancing the model's ICL ability.


\ding{175} \textbf{Providing the model with rich retrieved demonstrations would significantly improve its ICL ability.} To test this, we fine-tuned the model on a 4-shot instruction dataset, represented by the \textit{4-shot\_tuning} in Fig.~\ref{fig:performance}. The results indicate an improvement in the model's ICL ability. In 2-shot inference setting, comparing with \textit{0-shot\_tuning}, \textit{4-shot\_tuning} shows an average 8.2\% increase on classification tasks. However, the model's zero-shot generalization remained subpar on many tasks. We surmise that the model may learn shortcuts from the label words of the reference molecules rather than extracting the true relationships between molecular representations and properties.


\begin{figure*}
\centering
\includegraphics[width=1\linewidth]{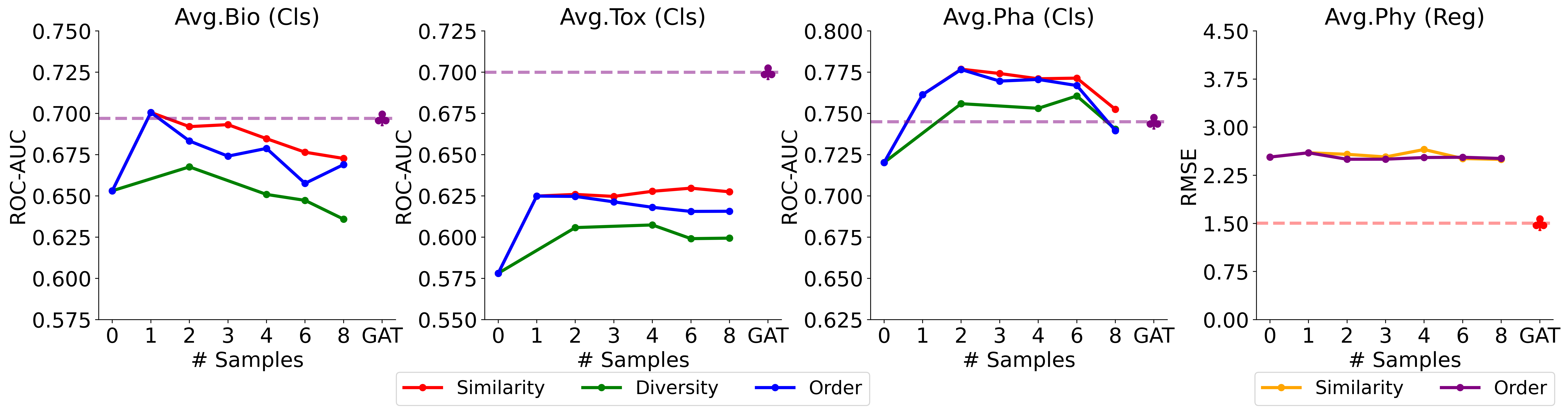}
\caption{The performance of MolecularGPT on Classifcation (Cls) and Regrassion (Reg) tasks with different in-context inference strategies. To show our model's remarkable capability, we also add the performance of the finetuned model, GAT.}
\label{fig:ICL}
\vspace{-0.4cm}
\end{figure*}

\ding{176} \textbf{Mixed-shot instruction sets are promising to optimize both zero-shot generalization and ICL abilities.} We developed two mixed instruction datasets: a combined \textit{0\&4-shot} and a comprehensive mix of $0,1,2,3,4$-shot \textit{(0-4-shot)} instruction datasets. As shown in \textit{0\&4-shot\_tuning} and \textit{0-4-shot\_tuning} in Fig.~\ref{fig:performance}, models fine-tuned on mixed-shot instruction datasets demonstrate a significant performance improvement compared to those fine-tuned on 0-shot or 4-shot instruction sets. This trend is consistently observed across various tested scenarios, like BACE, HIV, Tox21, BBBP and CYP450, indicating that our model derives the most benefit from mixed-shot instruction sets.

\ding{177} \textbf{Tuning on larger instruction set have exhibited superior performance across different tasks under both zero and few shot learning.} Models trained with larger datasets have exhibited superior performance on multi functional tasks, as evidenced by the improvements from GPT-2~\citep{gpt2} to GPT3~\citep{gpt3} and LLaMA2~\citep{llama} to LLaMA3~\citep{llama3}. To further enhance MolecularGPT, we double the size of the \textit{0\&4-shot} instruction sets. The results represented by the \textit{0\&4-shot\_tuning\_double} in Fig.~\ref{fig:performance} suggest that expanding the data scale enhances the model's performance across various tasks, like HIV, MUV and CYP450, either by zero-shot or few-shot learning.

\subsection{Hyperparameter Sensitivity Analysis}
\label{sub:icl}


To fully utilize the ICL ability of MolecularGPT, we now pay attention to the impact of question prompts and the retrieval examples to discuss the \textbf{RQ3}. Specifically, for question prompts, we test the robustness of our model respons to different instructions, shown in Tab.~\ref{table12}. As for few-shot demonstrations, we discuss the impact of molecular fingerprints in Tab.~\ref{table13}, as well as the number, order and diversity of demonstrations in Fig~\ref{fig:ICL}.



\ding{178} \textbf{The exceptional robustness of MolecularGPT is validated across different tasks.} Given the diversity and flexibility of natural language, we aim to evaluate the robustness of MolecularGPT against various instruction prompts. Adhering to the downstream datasets used in GIMLET, which provides five distinct types of instructions. We calculate the standard deviation of the ROC-AUC or RMSE metrics derived from these five instruction datasets. Comparative results with GIMLET is presented in Tab.~\ref{table12}. It is evident that our model demonstrates superior robustness compared to GIMLET across most tasks, with an average standard deviation of 0.06 compared to GIMLET's 0.08. This indicts the robustness of MolecularGPT that it genuinely comprehends complex instructions and can handle a range of property prediction tasks without requiring task-specific prompt designs.

\ding{179} \textbf{The effectiveness of MolecularGPT with different fingerprints for retrieving examples.} In both the 2-shot and 8-shot settings, our model demonstrates significant effectiveness when utilizing MACCS or MACCS-ECFP4 fingerprints for retrieving in-context examples, as illustrated in Tab.~\ref{table13}. The method based on MACCS-ECFP4 even outperforms our original MACCS-based method on 7 out of 10 tasks in the 8-shot setting. This indicates that our model exhibits strong structural awareness and generalization ability.

\begin{table}
\small  
\renewcommand\arraystretch{1.5}  
  \caption{Standard deviation for GIMLET and MolecularGPT in response to 5 types of instructions.}
  
  \centering
  \resizebox{\columnwidth}{!}{%
  \begin{tabular}{lcccccccccc}
    \toprule
    \multicolumn{1}{c}{}&\multicolumn{7}{c}{\textbf{Classification (AUC-ROC)}}&\multicolumn{3}{c}{\textbf{Regression (RMSE)}} \\
    \cmidrule(r){1-1}
    \cmidrule(r){2-8}
    \cmidrule(r){9-11}
    \textbf{Method} & \textbf{BACE} & \textbf{HIV} & \textbf{MUV} &  \textbf{Tox21} & \textbf{ToxCast}  & \textbf{BBBP} & \textbf{CYP450} & \textbf{ESOL} & \textbf{FreeSolv} & \textbf{Lipo}  \\
    \cmidrule(r){1-1}
    \cmidrule(r){2-8}
    \cmidrule(r){9-11}
    \textbf{GIMLET}& 0.024  & 0.034 & 0.017 & 0.009  & 0.005  & \textbf{0.013} & 0.012 & 0.020 & 0.532 & 0.082 \\
    \textbf{MolecularGPT}& \textbf{0.009}  & \textbf{0.018} & \textbf{0.012} & \textbf{0.008}  & \textbf{0.002}  & 0.045 & \textbf{0.002} & \textbf{0.007} & \textbf{0.498} &  \textbf{0.002} \\
    \bottomrule
  \end{tabular}
  }
  \label{table12}
\end{table}
    
\begin{table}
\small  
\renewcommand\arraystretch{1.5}  
  \caption{The performance of MolecularGPT when using different fingerprints for retrieving in-context examples.}
  
  \centering
  \resizebox{\columnwidth}{!}{%
  \begin{tabular}{lccccccccccc}
    \toprule
    \multicolumn{1}{c}{} & \multicolumn{1}{c}{} & \multicolumn{7}{c}{\textbf{Classification (AUC-ROC)}}&\multicolumn{3}{c}{\textbf{Regression (RMSE)}} \\
    \cmidrule(r){1-1}
    \cmidrule(r){2-2}
    \cmidrule(r){3-9}
    \cmidrule(r){10-12}
    \textbf{Method} & \textbf{Shot} & \textbf{BACE} & \textbf{HIV} & \textbf{MUV} &  \textbf{Tox21} & \textbf{ToxCast}  & \textbf{BBBP} & \textbf{CYP450} & \textbf{ESOL} & \textbf{FreeSolv} & \textbf{Lipo}  \\
    \cmidrule(r){1-1}
    \cmidrule(r){2-2}
    \cmidrule(r){3-9}
    \cmidrule(r){10-12}
    \textbf{MACCS} & \multirow{2}{*}{2-shot} & 0.7218 & \textbf{0.7204} & \textbf{0.6338} & 0.6573  & 0.5945 & 0.7260 & 0.8275 & \textbf{1.489} & 5.226 & \textbf{1.015} \\
    \textbf{MACCS-ECFP4}& & \textbf{0.7675}  & 0.6996 & 0.6042 & \textbf{0.6749} & \textbf{0.5993}  & \textbf{0.7298} & \textbf{0.8406} & 1.502  & \textbf{5.085} & 1.039 \\
    \cmidrule(r){1-1}
    \cmidrule(r){2-2}
    \cmidrule(r){3-9}
    \cmidrule(r){10-12}
    \textbf{MACCS} & \multirow{2}{*}{8-shot} & 0.7331 & 0.6382 & \textbf{0.6469} & 0.6565 & \textbf{0.5985} & 0.6822 & 0.8228 & \textbf{1.433} & 5.033 & 1.028 \\
    \textbf{MACCS-ECFP4}& & \textbf{0.7926} & \textbf{0.6674} & 0.6109 & \textbf{0.6685} & \textbf{0.5985} & \textbf{0.6962} & \textbf{0.8445} & 1.512 & \textbf{5.017} & \textbf{1.003} \\
    \bottomrule
  \end{tabular}
  }
  \label{table13}
\end{table}

\ding{180} \textbf{MolecularGPT gains significant enhancement with up to $2$ demonstrations, but the marginal benefit diminishes with additional retrieval molecules.} We investigate the impact of the number~\citep{ye2023context} of retrieval demonstrations, ranging from $0$ to $8$ examples based on similarity. The results indicate significant improvement when provided with up to $2$ examples cross all datasets. However, the performance does not get further improvement with more retrieval molecules. We hypothesize that: 1) More noise will be introduced with the increase of examples that has lower similarity with the query. 2) The maximum input length of $512$ tokens with at most $4$ examples in instructions constrains the model's capability while handling more examples.




\ding{181} \textbf{Ascending order of similarity for demonstrations is sub-optimal compared to descending order especially with more demonstrations.} We arrange the demonstrations~\citep{order, Calibrate} in a ascending order, placing the most similar examples at the end of k-shot instructions. The results in Fig.~\ref{fig:ICL} show that the ascending order is sub-optimal comparing to descending order, especially with more demonstrations which may be constrained by the model's long context capability. In an 8-shot setting, the ascending order shows an average performance that is 0.9\% lower than the descending order on classification tasks. We assume the model is more adaptable to reasoning with descending order by learning most related knowledge first.




\small{\textcircled{\scriptsize{\textbf{11}}}}\normalsize\;\textbf{Similar retrieved molecule demonstrations provides better performance than diverse demonstrations.} To increase the diversity, we retrieve equal number of molecules from each category~\citep{Fairness}. When the same number of examples is provided within instructions, the retrieval approach based on similarity consistently outperforms the one based on diversity across all classification tasks as shown in Fig.~\ref{fig:ICL}. In an 8-shot setting, the diversity-based method shows an average performance that is 2.7\% lower than the similarity-based on classification tasks.The similarity-based methodology tends to provide examples that align more coherently with the query molecules. In contrast, the diversity-based approach offers a mix of positive and negative examples, which potentially introduce noise and create ambiguity perplexing the language models.


\section{Conclusion}

In this study, we aim to equip the LLMs, particularly the LLaMA, with an expanded knowledge of molecular properties, enabling it to generalize to out-of-domain prediction tasks through zero-shot and few-shot ICL. We introduce MolecularGPT, a model that has been instruction tuned on over 1000 prediction tasks. Furthermore, we investigate the most effective types of instruction datasets for optimizing the model during both training and inference stages. Our findings demonstrate that MolecularGPT consistently outperforms baseline language models in few-shot scenarios and even surpasses supervised models on multiple datasets. In future work, we plan to incorporate additional molecular modalities and expand into other molecular-related tasks such as molecule captioning.

\section{Limitation}
In our research, we utilize SMILES strings to represent molecules. However, while effective, this approach overlooks the geometric structure information of real-world molecules, such as the 3D spatial position of each atom in a molecule. This limitation hinders our model's ability to represent molecular structures. Meanwhile, our work focuses solely on property prediction tasks and does not consider foundational tasks such as molecule optimization, molecule generation, and molecule captioning. This may restrict the potential applications of our model in practical settings. Lastly, although our model is compatible with supervised GNN models for classification tasks, we still have some gaps with them in regression tasks as directly generating numbers remains a challenge for nowadays foundational LLMs.

\bibliography{custom}

\clearpage
\appendix

\section{Datasets}

\subsection{Details of datasets}
\label{sub:Details of datasets}

We follow the dataset selection and splitting criteria specified in GIMLET~\citep{GIMLET}. As for downstream tasks, we divide the datasets in a ratio of 0.8:0.1:0.1 and report the results on the test sets. Moreover, considering the importance and extensive research in the field of quantum mechanical properties, we have included an additional two quantum mechanical properties: Highest Occupied Molecular Orbital(HOMO) and Lowest Unoccupied Molecular Orbital (LUMO), from the QM9 datasets~\citep{qm9} as our instruction tuning datasets. To construct the instructions for these additional datasets, we employed the method in Mol-Instructions~\citep{mol-instructions} and GIMLET~\citep{GIMLET}. Initially, we write a property description for each task according to Wikipedia and chemistry papers. Subsequently, we employ GPT-4.0~\citep{gpt4} to generate instructions based on these seed examples, resulting in various human question-framing styles instructions. The comprehensive list of tuning and downstream tasks are summarized in Tab.~\ref{table3}.

\begin{table*}
\caption{The overview of datasets}
\resizebox{\linewidth}{!}{
\begin{tabular}{lllccl}
\toprule
Splitting & Data Class & Dataset & No. of Molecules & No. of Tasks  & Task Type \\
    \midrule
     & Bioactivity assay & ChEMBL bioassay activity dataset & 365065 & 1048  & Classification \\
Tuning tasks& Physico-chemical & CHEMBL Property & 365065 & 13  & Regression \\
    & Quantum mechanical & QM9 & 267770 & 2  & Regression \\
    \midrule
    & Pharmacokinetic & CYP inhibition & 16896 & 5  & Classification \\
    & & BBBB Blood-brain barrier penetration & 2039 & 1  & Classification \\
    \cmidrule(r){2-6}
    & & MUV PubChem bioAssay & 93087 & 17  & Classification \\
    & Bio-activity & BACE-1 benchmark set & 1513 & 1  & Classification \\
    & & HIV replication inhibition & 41127 & 1  & Classification \\
    \cmidrule(r){2-6}
    Downstream tasks & Toxicity & Tox21Toxicology in the 21st century & 7831 & 12  & Classification \\
    & & Toxcast & 8598 & 617  & Classification \\
    \cmidrule(r){2-6}
    & & ESOL Water solubility & 1128 & 1  & Regression \\
    & Physico-chemical & FreeSolv Solvation free energy & 642 & 1 & Regression \\
    & & Lipo Lipophilicity & 4200 & 1  & Regression \\
    \bottomrule
    \end{tabular}}
  \label{table3}
\end{table*}

\subsection{Details of instructions}

Our instruction tuning datasets comprise three components: instruction, input, and output. The instruction component includes a description of the property along with some retrieval examples. The input is the SMILES string of the query molecule, while the output is the property label of query molecule. Here are a few examples of few-shot instructions from three tuning datasets: ChEMBL bioassay activity dataset,  CHEMBL Property dataset, and QM9 dataset.

A 1-shot instruction tuning sample from CHEMBL Property datasets:

\noindent
{\color{blue}"\#\#\# Instruction: Aromatic rings (also known as aromatic compounds or arenes) are hydrocarbons which contain benzene, or some other related ring structure. Here are some examples. \\
SMILES: Cc1ccc2ccccc2n1 \\
label: 2 \\
Please count how many aromatic rings exist in this molecule. \\
\#\#\# Input: Cc1ccnc2ccccc12 \\
\#\#\# Response: 2"} \\

A 3-shot instruction tuning sample from ChEMBL bioassay activity datasets:

\noindent
{\color{blue}"\#\#\# Instruction: The assay is PUBCHEM\_BIOASSAY: NCI human tumor cell line growth inhibition assay. Data for the DMS 273 Small Cell Lung cell line. (Class of assay: confirmatory), and it is Target assigned is non-molecular. The assay has properties: assay category is confirmatory; assay cell type is DMS-273; assay type description is Functional. Here are some examples. \\
SMILES: CC(C)C(N)=O \\
label: No \\
SMILES: O=CNC=Cc1ccccc1 \\
label: No \\
SMILES: COC(=O)C\#CC(N)=O \\
label: No \\
Is the molecule effective to this assay? \\
\#\#\# Input: CNC=O \\
\#\#\# Response: No"} \\

A 4-shot instruction tuning sample from QM9 datasets:

\noindent
{\color{blue}"\#\#\# Instruction: Lumo is the Lowest unoccupied molecular orbital energy. Here are some examples. \\
SMILES: CC \\
label: 0.1 \\
SMILES: CC(C\#C)C\#CC\#C \\
label: -0.02 \\
SMILES: CC\#CC\#CC\#C  \\
label: -0.05 \\
SMILES: CC(C\#C)C\#C \\
label: 0.03  \\
What is Lumo value of this molecule? \\
\#\#\# Input: C1CC1 \\
\#\#\# Response: 0.1"} 

\section{Training Setup}

\begin{table*}
\scriptsize
\renewcommand\arraystretch{0.5}
  \caption{The zero-shot inference results under different types of instructions: the original, detailed, expanded, rewritten, and shortened instructions.}
  
  \centering
  \begin{tabular}{ccccccccccc}
    \toprule
    \multicolumn{1}{c}{}&\multicolumn{7}{c}{Classification (AUC-ROC)}&\multicolumn{3}{c}{Regression (RMSE)} \\\\
    \cmidrule(r){1-1}
    \cmidrule(r){2-8}
    \cmidrule(r){9-11}
    Instruction type  & BACE & HIV & MUV &  Tox21 & ToxCast  & BBBP & CYP450 & ESOL & FreeSolv & Lipo  \\\\
    \cmidrule(r){1-1}
    \cmidrule(r){2-8}
    \cmidrule(r){9-11}
    Original& 0.6212 & 0.7128 & 0.6253 &  0.5893 & 0.5669 & 0.6373 & 0.8031 & 1.471 & 4.975 & 1.157 \\\\
    Detailed& 0.6222 & 0.6754 & 0.6090 &  0.6047 & 0.5710 & 0.6600 & 0.8076 & 1.457 & 5.036 & 1.158 \\\\
    Expanded& 0.6175 & 0.7134 & 0.6017 &  0.6110 & 0.5688 & 0.6511 & 0.8053 & 1.474 & 5.023 & 1.154 \\\\
    Rewritten& 0.6351 & 0.6893 & 0.6172 & 0.5955  & 0.5666  & 0.6427 & 0.8050 & 1.457 & 5.018 & 1.157 \\\\
    Shortened& 0.6409  & 0.6697 & 0.6348 & 0.5924  & 0.5692  & 0.5374 & 0.8032 & 1.462 & 6.258 & 1.158 \\\\
    \cmidrule(r){1-1}
    \cmidrule(r){2-8}
    \cmidrule(r){9-11}
    Standard deviation& 0.0090  & 0.0183 & 0.0117 & 0.0081  & 0.0016  & 0.0448 & 0.0016 & 0.0071 & 0.4984 &  0.0015 \\\\
    \bottomrule
  \end{tabular}
  \label{table9}
\end{table*}

\begin{table*}
\scriptsize
\renewcommand\arraystretch{0.5}
  \caption{The zero- and few-shot performances of model which was fine-tuned on 0-shot instruction datasets.}
  
  \centering
  \begin{tabular}{lccccccccccc}
    \toprule
    \multicolumn{2}{c}{Tasks}&\multicolumn{7}{c}{Classification (AUC-ROC)}&\multicolumn{3}{c}{Regression (RMSE)} \\\\
    \cmidrule(r){1-2}
    \cmidrule(r){3-9}
    \cmidrule(r){10-12}
    Method & Type & BACE & HIV & MUV &  Tox21 & ToxCast  & BBBP & CYP450 & ESOL & FreeSolv & Lipo  \\\\
    \cmidrule(r){1-2}
    \cmidrule(r){3-9}
    \cmidrule(r){10-12}
    & 0-Shot & 0.6033 & 0.6028 & 0.6010 & 0.5824  & 0.5839  & 0.6521 & 0.7684 & 1.767 & 5.185 & 1.163 \\\\
    & 1-Shot & 0.6297 & 0.4671 & 0.5740 & 0.6016  & 0.5886  & 0.6436 & 0.7667 & 1.442 & 5.324 & 1.032 \\\\
    & 2-Shot & 0.5903 & 0.4006 & 0.5665 & 0.5956  & 0.5867  & 0.6166 & 0.7556 & 1.438 & 5.482 & 1.053 \\\\
    0\_examples& 3-Shot& 0.5344 & 0.4151 &  0.5705 &  0.5974 & 0.5757 & 0.6032 & 0.7457 & 1.379 & 5.617 & 1.016 \\\\
    & 4-Shot & 0.5334 & 0.4393 & 0.5675 & 0.5942  &  0.5828 & 0.6197 & 0.7367 & 1.249 & 5.555 & 1.010 \\\\
    & 6-Shot & 0.5314 & 0.3784 & 0.5312 &  0.5843 &  0.5723 & 0.5767 & 0.7374 & 1.241 & 5.961 & 0.979 \\\\
    & 8-Shot & 0.4388 & 0.3768 & 0.5637 & 0.5724  &  0.5672 & 0.5187 & 0.7050 & 1.131 & 5.852 & 0.984 \\\\
    \bottomrule
  \end{tabular}
  \label{table4}
\end{table*}

\begin{table*}
\scriptsize
\renewcommand\arraystretch{0.5}
  \caption{The zero- and few-shot performances of model which was fine-tuned on 4-shot instruction datasets.}
  
  \centering
  \begin{tabular}{lccccccccccc}
    \toprule
    \multicolumn{2}{c}{Tasks}&\multicolumn{7}{c}{Classification (AUC-ROC)}&\multicolumn{3}{c}{Regression (RMSE)} \\\\
    \cmidrule(r){1-2}
    \cmidrule(r){3-9}
    \cmidrule(r){10-12}
    Method & Type & BACE & HIV & MUV &  Tox21 & ToxCast  & BBBP & CYP450 & ESOL & FreeSolv & Lipo  \\\\
    \cmidrule(r){1-2}
    \cmidrule(r){3-9}
    \cmidrule(r){10-12}
    & 0-Shot & 0.5446 & 0.5514 & 0.6406 & 0.5425  & 0.5588  & 0.4709 & 0.6282 & 2.703 & 4.620 & 1.144 \\\\
    & 1-Shot & 0.6773 & 0.5135 & 0.6240 & 0.6911  & 0.6140  & 0.6342 & 0.8239 & 1.644 & 5.062 & 1.019 \\\\
    & 2-Shot & 0.6860 & 0.5626 & 0.6203 & 0.7053  & 0.6163  & 0.6563 & 0.8420 & 1.278 & 4.942 & 0.949 \\\\
    4\_examples& 3-Shot  & 0.7315 & 0.5577 & 0.6269  & 0.7096  & 0.6220 & 0.6533 & 0.8479  & 1.277 & 4.734 & 0.949 \\\\
    & 4-Shot & 0.7264 & 0.5624 & 0.6238 & 0.7233  & 0.6243  & 0.6644 & 0.8525 & 1.311 & 4.978 & 0.956 \\\\
    & 6-Shot & 0.7294 & 0.5768 & 0.6115 & 0.7339  & 0.6268  & 0.6553 & 0.8523 & 1.284 & 4.941 & 0.974 \\\\
    & 8-Shot & 0.7327 & 0.6234 & 0.6079 & 0.7396  & 0.6271  & 0.6430 & 0.8554 & 1.254 & 4.889 & 0.967 \\\\
    \bottomrule
  \end{tabular}
  \label{table5}
\end{table*}

To efficiently finetune the LLaMA2-chat-7B, we employed QLoRA \citep{qlora} approach. To enhance memory utilization and speed up the training process, we incorporated Deepspeed ZeRO stage 2 \citep{deepspeed}, FlashAttention-2 \citep{flash2}, and BFloat16 mixed precision techniques. We set the learning rate to 3e-4 and the maximum inputs length to 512 tokens. All models were trained on 4 Tesla A800-80G GPUs and inferenced on 1 RTX 3090 GPU.

\section{Detailed Experiment Results}
\subsection{The robustness of MolecularGPT}

To evaluate the robustness of MolecularGPT across diverse instructional phrasings, we adopt the instruction datasets constructed in GIMLET~\citep{GIMLET}, which utilizes GPT-3.5-turbo to generate four distinct types of instructions based on the original instruction: detailed, expanded, rewritten, and shortened instructions. We present the zero-shot inference results derived from these diverse instructions and compute their ROC-AUC or RMSE standard deviation, as outlined in Tab.~\ref{table9}. Our findings suggest that MolecularGPT exhibits robust performance across different instructional variations.

\subsection{The effect of instruction datasets}
To find a model with superior zero-shot generalization and ICL capabilities, we assess the performance of models that have been fine-tuned by datasets that employ diverse mixture strategies. These strategies include single 0-shot instruction, single 4-shot instruction, combined 0\&4-shot instruction, combined 0,1,2,3,4-shot (0-4 shot) instruction, and doubled scale of combined 0\&4-shot instruction datasets.

\begin{table*}
\scriptsize
\renewcommand\arraystretch{0.5}
  \caption{The zero- and few-shot performances of model which was fine-tuned on 0\&4-shot instruction datasets.}
  
  \centering
  \begin{tabular}{lccccccccccc}
    \toprule
    \multicolumn{2}{c}{Tasks}&\multicolumn{7}{c}{Classification (AUC-ROC)}&\multicolumn{3}{c}{Regression (RMSE)} \\\\
    \cmidrule(r){1-2}
    \cmidrule(r){3-9}
    \cmidrule(r){10-12}
    Method & Type & BACE & HIV & MUV &  Tox21 & ToxCast  & BBBP & CYP450 & ESOL & FreeSolv & Lipo  \\\\
    \cmidrule(r){1-2}
    \cmidrule(r){3-9}
    \cmidrule(r){10-12}
    & 0-Shot & 0.6568 & 0.6728 & 0.5533 &  0.6067 & 0.5352  & 0.6086 & 0.7931 & 1.377 & 5.376 & 1.208 \\\\
    & 1-Shot & 0.7393 & 0.6620 & 0.5954 & 0.6817  & 0.5809  & 0.7087 & 0.8231 & 1.468 & 5.034 & 1.042  \\\\
    & 2-Shot & 0.7204 & 0.6485 & 0.5969 & 0.7004  & 0.5863  & 0.7135 & 0.8357 & 1.481 & 4.981 & 1.038 \\\\
    0,4\_examples& 3-Shot  & 0.7543 & 0.6459 & 0.6139  & 0.6964 & 0.5877 & 0.6997 & 0.8368 & 1.481 & 4.984 & 1.030 \\\\
    & 4-Shot & 0.7593 & 0.6363 & 0.6026 & 0.7074  & 0.5938  & 0.7130 & 0.8390 & 1.413 & 5.149 & 1.028 \\\\
    & 6-Shot & 0.7574 & 0.6150 & 0.5926 & 0.7156  & 0.5954  & 0.7145 & 0.8438  & 1.427 & 4.928 &  1.047 \\\\
    & 8-Shot & 0.7474 & 0.6197 & 0.5942 & 0.7182  &  0.5962 & 0.7029 & 0.8459 & 1.479 & 4.846  & 1.031 \\\\
    \bottomrule
  \end{tabular}
  \label{table6}
\end{table*}

\begin{table*}
\scriptsize
\renewcommand\arraystretch{0.5}
  \caption{The zero- and few-shot performances of model which was fine-tuned on 0,1,2,3,4-shot instruction datasets.}
  
  \centering
  \begin{tabular}{lccccccccccc}
    \toprule
    \multicolumn{2}{c}{Tasks}&\multicolumn{7}{c}{Classification (AUC-ROC)}&\multicolumn{3}{c}{Regression (RMSE)} \\\\
    \cmidrule(r){1-2}
    \cmidrule(r){3-9}
    \cmidrule(r){10-12}
    Method & Type & BACE & HIV & MUV &  Tox21 & ToxCast  & BBBP & CYP450 & ESOL & FreeSolv & Lipo  \\\\
    \cmidrule(r){1-2}
    \cmidrule(r){3-9}
    \cmidrule(r){10-12}
    & 0-Shot & 0.6521 & 0.7046 & 0.5788 & 0.5673  & 0.5612  & 0.6807 & 0.7539 & 1.228 & 5.835 & 1.176  \\\\
    & 1-Shot & 0.7728 & 0.7049 & 0.5859 &  0.6639 &  0.6026 & 0.7220 & 0.8115 & 1.192 & 4.979 & 0.996 \\\\
    & 2-Shot & 0.7393 & 0.6816 & 0.5866 &  0.6780 & 0.6085 & 0.7360 & 0.8232 & 1.218 & 4.985 & 0.983 \\\\
    0-4\_examples& 3-Shot  & 0.7793 & 0.6806 & 0.5993  & 0.6719  & 0.6066 & 0.7187 & 0.8323 & 1.223 & 4.979 & 0.960 \\\\
    & 4-Shot & 0.7743 & 0.6807 & 0.5849 &  0.6817 & 0.6148  & 0.7272 & 0.8394 & 1.167 & 5.247 & 0.983 \\\\
    & 6-Shot & 0.7724 & 0.6673 & 0.6044 &  0.6956 & 0.6179  & 0.7223 & 0.8452 & 1.165 & 5.219 & 0.976 \\\\
    & 8-Shot & 0.8102 & 0.6724 & 0.6170 &  0.7043 & 0.6190  & 0.7125 & 0.8418 & 1.163 & 5.033 & 0.992 \\\\
    \bottomrule
  \end{tabular}
  \label{table7}
\end{table*}

\begin{table*}
\scriptsize
\renewcommand\arraystretch{0.5}
  \caption{The zero- and few-shot performances of model which was fine-tuned on double scale 0\&4-shot instruction datasets.}
  
  \centering
  \begin{tabular}{lccccccccccc}
    \toprule
    \multicolumn{2}{c}{Tasks}&\multicolumn{7}{c}{Classification (AUC-ROC)}&\multicolumn{3}{c}{Regression (RMSE)} \\\\
    \cmidrule(r){1-2}
    \cmidrule(r){3-9}
    \cmidrule(r){10-12}
    Method & Type & BACE & HIV & MUV &  Tox21 & ToxCast  & BBBP & CYP450 & ESOL & FreeSolv & Lipo  \\\\
    \cmidrule(r){1-2}
    \cmidrule(r){3-9}
    \cmidrule(r){10-12}
    & 0-Shot & 0.6212 & 0.7128 & 0.6253 & 0.5893  & 0.5669  & 0.6373 & 0.8031 & 1.471 & 4.975 &  1.157\\\\
    & 1-Shot & 0.7520 & 0.7172 & 0.6327 & 0.6529 & 0.5968  & 0.6999 & 0.8229 & 1.496 & 5.248 & 1.058 \\\\
    & 2-Shot & 0.7218 & 0.7204 & 0.6338 & 0.6573  & 0.5945  & 0.7260 & 0.8275 & 1.489  & 5.226 & 1.015 \\\\
    0,4\_examples\_double& 3-Shot  & 0.7350 & 0.7038 & 0.6408  &  0.6542 & 0.5951 & 0.7191 & 0.8293 & 1.494 & 5.082 & 1.032 \\\\
    & 4-Shot & 0.7228 & 0.6893 & 0.6419 & 0.6577  &  0.5978 & 0.7168 & 0.8252 & 1.535  & 5.375 & 1.045 \\\\
    & 6-Shot & 0.7181 & 0.6554 & 0.6561 &  0.6629 &  0.5965 & 0.7139 & 0.8289 & 1.465 & 5.046 & 1.023 \\\\
    & 8-Shot & 0.7331 & 0.6382 & 0.6469 &  0.6565 & 0.5985  & 0.6822 & 0.8228 & 1.433 & 5.033 & 1.028 \\\\
    \bottomrule
  \end{tabular}
  \label{table8}
\end{table*}

\begin{table*}
\scriptsize
\renewcommand\arraystretch{0.5}
  \caption{The few-shot inference results of MolecularGPT using a ICL template that organizes the retrieval demonstrations in a ascending order.}
  
  \centering
  \begin{tabular}{ccccccccccc}
    \toprule
    \multicolumn{1}{c}{}&\multicolumn{7}{c}{Classification (AUC-ROC)}&\multicolumn{3}{c}{Regression (RMSE)} \\\\
    \cmidrule(r){1-1}
    \cmidrule(r){2-8}
    \cmidrule(r){9-11}
    Type  & BACE & HIV & MUV &  Tox21 & ToxCast  & BBBP & CYP450 & ESOL & FreeSolv & Lipo  \\\\
    \cmidrule(r){1-1}
    \cmidrule(r){2-8}
    \cmidrule(r){9-11}
    2-shot& 0.7105 & 0.7126 & 0.6269 &  0.6553 & 0.5941 & 0.7245 & 0.8287 & 1.514 & 4.934 & 1.053 \\\\
    3-shot& 0.7172 & 0.6884 & 0.6166 &  0.6489 & 0.5938 & 0.7090 & 0.8302 & 1.527 & 4.898 & 1.078 \\\\
    4-shot& 0.7333 & 0.6732 & 0.6299 &  0.6474 & 0.5888 & 0.7130 & 0.8281 & 1.500 & 5.031 & 1.050 \\\\
    6-shot& 0.7067 & 0.6423 & 0.6237 &  0.6447 & 0.5864 & 0.7040 & 0.8297 & 1.446 & 5.097 & 1.049 \\\\
    8-shot& 0.7407 & 0.6311 & 0.6352 &  0.6452 & 0.5861 & 0.6555 & 0.8237 & 1.462 & 5.041 & 1.034 \\\\
    \bottomrule
  \end{tabular}
  \label{table10}
\end{table*}

\begin{table*}
\scriptsize
\renewcommand\arraystretch{0.5}
  \caption{The few-shot inference results of MolecularGPT, which retrieves demonstrations based on their diversity.}
  
  \centering
  \begin{tabular}{cccccccc}
    \toprule
    \multicolumn{1}{c}{}&\multicolumn{7}{c}{Classification (AUC-ROC)} \\\\
    \cmidrule(r){1-1}
    \cmidrule(r){2-8}
    Type  & BACE & HIV & MUV &  Tox21 & ToxCast  & BBBP & CYP450   \\\\
    \cmidrule(r){1-1}
    \cmidrule(r){2-8}
    2-shot& 0.7039 & 0.6854 & 0.6135 &  0.6297 & 0.5819 & 0.7037 & 0.8081  \\\\
    4-shot& 0.6688 & 0.6584 & 0.6255 &  0.6321 & 0.5826 & 0.6962 & 0.8100  \\\\
    6-shot& 0.6782 & 0.6425 & 0.6213 &  0.6184 & 0.5797 & 0.7079 & 0.8133  \\\\
    8-shot& 0.6832 & 0.6127 & 0.6118 &  0.6140 & 0.5848 & 0.6740 & 0.8070  \\\\
    \bottomrule
  \end{tabular}
  \label{table11}
\end{table*}
In the combined 0\&4-shot methodology, we merge the 0-shot and 4-shot instruction datasets in an equal ratio of 0.5: 0.5. For the comprehensive 0-4 shot mix, we integrate the 0,1,2,3, and 4-shot instruction datasets in a ratio of 0.6: 0.1: 0.1: 0.1: 0.1. During these procedures, we ensure the absence of duplicate query molecules and maintain the scale of the datasets. For the doubled scale of 0\&4-shot, we amalgamate the 0-shot and 4-shot instruction datasets in an equal proportion of 1: 1. The results of the zero- and few-shot inferences are presented in the following Tab.~\ref{table4},~\ref{table5},~\ref{table6},~\ref{table7} and ~\ref{table8}.


\subsection{The effect of inference strategies}

We examine the efficacy of the order of the demonstrations within instructions. Tab.~\ref{table10} illustrates the performance of arranging retrieval demonstrations in ascending order. Notably, the phrasing in zero-shot or one-shot instruction is consistent in both ascending and descending order. Consequently, we present the results of 2-shot and above.
Additionally, we examine the efficacy of retrieval based on diversity, comparing it with a strategy that prioritizes similarity, as illustrated in Tab.~\ref{table11}. It's important to note that to ensure an equal distribution of different class samples, evaluating even-numbered shot is essential. Moreover, this strategy is specifically designed for classification tasks, as regression tasks lack distinct classes.

\end{document}